\documentstyle [12pt] {article}

\catcode`@=12
\newcommand{\be}{\begin{equation}}
\newcommand{\ee}{\end{equation}}
\newcommand{\ber}{\begin{eqnarray}}
\newcommand{\eer}{\end{eqnarray}}
\newcommand{\bers}{\begin{eqnarray*}}
\newcommand{\eers}{\end{eqnarray*}}
\begin{document}
\vspace{0.5in}
\oddsidemargin -.375in
\def\ee{\end{equation}}
\thispagestyle{empty}
\begin{flushright} ISU-NP-97-09\\UTPT-97-23\\November 1997\
\end{flushright}
\vspace {.5in}
\begin{center}
{\Large\bf Decay constants, semi-leptonic and non-leptonic\newline
 $B                    $ decays in a Bethe-Salpeter Model 
% $B \rightarrow D(D^*)$ decays in a Bethe-Salpeter Model 
 \\}
\vspace{.5in}
{\bf A. Abd El-Hady${}^a$, 
Alakabha Datta ${}^b$ and J.~P.~Vary ${}^a$\\}
\vspace{.1in}
${}^{a)}$  {\it
 Department of Physics and Astronomy, Iowa State University , Ames, Iowa
50011, USA.}\\
\vspace{.1in}
${}^{b)}$  {\it
Department of Physics, University of Toronto, Toronto, Ontario, Canada
M5S 1A7
} \\ 

\vskip .5in
%{\bf Abstract}
\end{center}

\vskip .1in \begin{abstract} We evaluate the decay constants for the $B$
and $D$ mesons and the form factors for the
semileptonic decays of the $B$ meson to $D$ and $D^*$ mesons in a
Bethe-Salpeter model.
From data we extract $V_{cb}=0.039 \pm 0.002$ from
${\bar B} \to D^* l {\bar{\nu}} $ and $V_{cb}=0.037 \pm 0.004$ from
${\bar B} \to D l {\bar{\nu}}  $  decays.
 The form factors are then used to obtain
non-leptonic decay partial widths for $ B\to D \pi (K)  $ and $B \to D D (D_s) $ in the
%non-leptonic decay partial widths for $ B\to D K (\pi) $ and $B \to D D $ in the
factorization approximation. 

\end{abstract}
\vskip .25in
\newpage

\section{INTRODUCTION}
In previous papers \cite {BS,BSsep} we have developed a model for mesons
based on the Bethe-Salpeter equation (BSE). Recently \cite {BSV,BSV1}, we
calculated the form factors in the semileptonic $ B \to D (D ^*) l \nu $
decays and extracted the  
Cabibbo-Kobayashi-Maskawa (CKM) 
matrix
element
$V_{cb}$ from data. The key ingredient in the computation of the form
factor was the construction of the physical states for the $B$ and $D$
mesons in terms of the wavefunction obtained by solving a reduced BSE.
 In this paper we improve upon the work of Ref.\cite
{BSV1} in two ways. First we establish a theoretical connection between the matrix element of the bare current operator
calculated using the model states constructed in
\cite {BSV1} and the  matrix elements of an effective current operator based upon arguments
for contributions from neglected configurations.
Second, we make an ansatz for the
correspondence between the matrix elements of the bare and the effective current operator.
 The effective current operators 
 are then used to calculate not only the decay constants and semileptonic form
factors similar to our previous work but also the branching fractions
 of non-leptonic decays. 

The discovery of Heavy Quark Symmetry (HQS) in recent years
\cite{H4,H5,H6,H7,H71} has generated considerable interest in the study
of systems containing heavy quark(s). It has been shown that, in the
heavy quark limit, the properties of systems containing a heavy quark
are greatly simplified. HQS results in relations between
non-perturbative quantities, such as form factors, for different
processes involving transitions of a heavy quark to another quark. The
development of Heavy Quark Effective Theory (HQET) \cite {H6} allows one
to systematically calculate corrections to the results of the HQS limit
in inverse powers of the heavy quark mass $m_Q$. 
In spite of impressive results obtained in HQET, it has not solved the
problem of calculating the transition form factors in QCD. In
particular, HQS reveals relations between form factors but does not
provide a determination of the form factors themselves. Furthermore, the systematic
expansion of the form factors in $1/m_Q$ in HQET involves additional
non-perturbative matrix elements which are not calculable from first
principles. 

We are thus
forced to rely on models for the non-perturbative quantities. However,
the constraints of HQET, which are based on QCD, allow one to construct
models which are consistent with HQET and hence QCD. We have already
demonstrated the consistency of our model with the requirements of HQET 
\cite{BSV1}.

The parameters in the BSE
  are fixed by fitting the meson spectrum. Hadronic states necessary for
the calculation of form factors are constructed with the BS
wavefunctions. In our formalism the mesons have been considered as composed of
 $q{\bar{q}}$ constituent quarks which defines the limits of our model
space. Our dual thrust in this effort is to correct for the limited
model space and to carry out new applications. Higher Fock state effects are introduced
 though an ansatz, involving an additional parameter, connecting the bare current operator to an effective
operator. 

 The additional parameter introduced in this process
is chosen by a fit to selected experimental data. An evaluation of
the 
  decay constants of the B and D system along
with the semi-leptonic form factors and the non-leptonic decays is performed
 without any additional free
parameter and are, therefore, viewed as
predictions of this model. Based on this new approach we again extract $V_{cb}$ from the measured
differential decay rate of ${\bar {B}}\rightarrow D^* l {\bar{\nu}}$ and find a $20\%$
increase over our previous results \cite{BSV1}. We also extract $V_{cb}$
using recent measurement of ${\bar B} \to D l {\bar{\nu}} $ by CLEO
\cite{newcleo}. The two resulting values for $V_{cb}$ presented in this
paper are consistent with each other.

The paper is organized as follows:  In Section 2 we give a brief
review of
Bethe-Salpeter model for mesons. In Section 3, we discuss the
formalism for the calculation of the decay constants and the
 form factors after establishing the connection between the bare current oprator and the
effective current operator.
 In Section 4, we discuss non-leptonic decays and in Section 5 we
present and discuss the results of our work.

\section{ BETHE-SALPETER MODEL FOR MESONS}

{ In Ref.\cite{BS} we have developed a model for mesons 
based on the Bethe-Salpeter
equation. The wavefunctions for the mesons were solved from
 three dimensional reductions
of BSE, called the Quasi-potential equations (QPE). It was found that
two reductions give a good description of the meson spectrum, including
open flavour mesons, over a wide range of states. Masses for 47 states 
were predicted using
seven parameters given below with mass root mean square deviation of about 50
MeV Ref.\cite{BSsep}.

The interaction kernel in the BSE is written as a sum of a one-gluon exchange
interaction in the ladder approximation, $V_{OGE}$, and a
phenomenological, long-range linear confinement potential, $V_{CON}$.  
In momentum-space this interaction takes the form,
\begin{eqnarray} V_{OGE}+V_{CON} & = & {4\over
3}\alpha_s{\gamma_\mu\otimes\gamma_\mu\over {(q-q')^2}}
+\sigma{\rm\lim_{\mu\to 0}}{\partial^2\over\partial\mu^2} {{\bf
1}\otimes{\bf 1}\over-(q-q')^2+\mu^2} \ \end{eqnarray}
Here, $\alpha_s$ is the strong coupling, which is weighted by the meson
color factor of ${4\over 3}$, and the string tension $\sigma$ is the
strength of the confining part of the interaction.  We adopt a scalar
Lorentz structure $V_{CON}$ as discussed in \cite {BSsep}

In our model 
the strong coupling is assumed  to run as in the leading
log expression for $\alpha_s$,
\begin{eqnarray} \alpha_s(Q^2) & = & {4\pi\alpha_s(\mu^2)\over
4\pi+\beta_1\alpha_s(\mu^2){\rm ln}\bigl({Q^2/\mu^2}\bigr)} \
\end{eqnarray} 
where $\beta_1=11-2n_f/3$ and $n_f$ is the number of quark flavors, with
$\alpha_s(\mu^2=M_Z^2)\simeq 0.12$ where
 $Q^2$ is related to to the meson mass scale through,
\begin{equation}
Q^2 = \gamma^2 M_{meson}^2 + \beta^2,
\end{equation}
where $\gamma$ and $\beta$ are parameters determined by a fit to the meson
spectrum.

In our formulation of BSE there are therefore seven parameters : four masses,
$m_{u}$=$m_{d}$, $m_c$, $m_{s}$, $m_{b}$; the string tension $\sigma$,
and the parameters $\gamma$ and $\beta$ used to govern the running of
the coupling constant. Once the parameters are fixed from the mass
spectrum, 
the meson
wavefunctions from the BSE can be used to predict physical
observables. 

Table 1 shows the values of the parameters used in two
reductions of Bethe-Salpeter equation referred to as A, B reductions
\cite {BSsep}.

\begin{table}[tbh]
\caption{Values of the parameters used in reductions A,B together with root mean square deviation from experimental meson masses}
\begin{center}
\begin{tabular}{|c|c|c|}
\hline
   &Reduction A   &  Reduction B  \\
\hline
$m_b$ (GeV) &4.65&4.68\\
$m_c$ (GeV) &1.37&1.39\\
$m_s$ (GeV) &0.397&0.405\\
$m_u$ (GeV) &0.339&0.346\\
$\sigma$ (GeV$^2$) &0.233&0.211\\
$\gamma$  &0.616&0.444\\
$\beta$ (GeV)  &0.198&0.187\\
RMS (MeV)&43&50\\
\hline
\end{tabular}\\
\end{center}
\end{table}

\section{ Decay Constants and Semi-Leptonic form factors}
The weak decay constants for the heavy hadrons are defined below
\ber
<0|J_{\mu}|P(p)> & = & i f_P p_{\mu}\nonumber\\
<0|J_{\mu}|V(p)> & = & m_Vf_V \varepsilon_{\mu}\nonumber\\
J_{\mu} & = & V_{\mu} -A_{\mu}\
\eer
where $P$ and $V$ are pseudo-scalar and vector states and $V_{\mu}$ and
$A_{\mu}$ are the vector and axial vector currents.

The  Lagrangian for the semileptonic decays involving the $b \rightarrow c$
 transition has the standard current-current form after the $W$ boson is
integrated out in the effective theory.  \begin{eqnarray} H_{W} & = &
\frac{G_F}{2 \sqrt{2}}V_{cb} {\bar c}\gamma_{\mu}(1-\gamma_{5}) b
{\bar{\nu}}\gamma^{\mu}(1-\gamma_{5})l \ \end{eqnarray} The leptonic
current in the effective interaction is completely known and the matrix
element of the vector ($V_{\mu}$) and the axial vector ($ A_{\mu}$)
hadronic currents between the meson states are represented in terms of
form factors which are defined in the equations below \cite{BSW}.  

\ber
	<D(p_D)|J_{\mu} | B(p_B) >
	&=& \left[ (p_B+ p_D)_\mu - \frac{m_B^2-m_D^2}{q^2} q_\mu \right]
	  F_1(q^2)\nonumber\\
	 & + & \frac{m_B^2-m_D^2}{q^2} q_\mu F_0(q^2) \
\eer
where $q = p_B -p_D $.

\ber
<D^{*}(p')|J^{\mu}|B(p)>
& = & b_0 \varepsilon^{\mu \nu \alpha
\beta}\varepsilon_{\nu}^{*}p_{\alpha}
p'_{\beta}
+ b_1 \varepsilon ^{*\mu} + b_2(p+p')^{\mu}
+b_3(k)^{\mu}\
\eer
with
\bers
	b_0 & = & \frac{2V(k^2)}{m_B + m_{D^*}} \\
	b_1 & = & i(m_B + m_{D^*}) A_1(k^2) \\
	b_2 & = &-i \varepsilon^{*} \cdot k
 \frac{A_2(k^2)}{m_B + m_{D^*}} \\
	b_3 & = & 
i \varepsilon^{*} \cdot k \frac{2m_{D^*} (A_0(k^2) - A_3(k^2))}{k^2}\\
        A_3(k^2) &=& \frac{(m_B + m_{D^*})A_1(k^2) - (m_B - m_{D^*})A_2(k^2)}
{2m_{D^*}}\
\eers
where $k=p_B-p_{D^*}$.
$F_0, F_1,
V, A_0, A_1$, $A_2$, and $A_3$ are Lorentz invariant form factors which are scalar
functions of the momentum transfer $  (P_B -P_D(P_{D^*}))^2$.
 The calculation of the decay constants and
 form factors proceeds in two steps. In the
first step, the full current from QCD is matched to the current in the
effective theory (HQET) at the heavy quark mass scale \cite{NR}.
 Renormalization group equations are then used to run down to
a low energy scale $\mu \sim 1 $ GeV where the constraints of HQET
operate and where it is reasonable to calculate matrix elements in a
valence constituent quark model like the one we employ here \cite{ISGW2}.
We have already described the first step in our previous publication
\cite{BSV1} and therefore we will not repeat it here.

 The second step is the calculation of the matrix
elements of the currents in the model to obtain the decay constants and
form factors .
 Such a
        calculation requires the knowledge of the
         meson wavefunctions. In our formalism the mesons
        are taken as bound states of a quark and an antiquark. The wavefunctions 
for the mesons, as already mentioned, are
calculated  by solving the Bethe-Salpeter equation \cite
{BS,BSsep}. 
We construct
  the meson states as \cite{ISGW1}
\begin{eqnarray}
|M({\bf {P_M}},J,m_J)\rangle\  & = &
\sqrt{2M_H} \int d^{3}{\bf p} \langle L m_{L}S m_{S}|J m_J\rangle\  \langle
s m_s \bar{s} m_{\bar{s}}|S m_S\rangle\ \nonumber\\
 & &\Phi_{L m_L}({\bf p})|\bar q( {m_{\bar
q} \over \ M } {\bf {P}_M} - {\bf {p}},m_{\bar s})
\rangle|q({m_q \over \ M } {\bf {P}_M} + {\bf {p}},m_s)\rangle
\end{eqnarray}
where

\begin{eqnarray}
|q({\bf {p}},m_s)\rangle\ &=& \sqrt{\frac{(E_q + m_q)}{2m_q}} \pmatrix{
 \chi^{m_s} \cr \ {{\bf{\sigma}}\cdot{\bf p }\over
{(E_q+m_q)}}  \chi^{m_s} \cr }\nonumber\\
M&=&m_q+m_{\bar q}\nonumber\\ 
E_q&=&\sqrt{m^2_{q}+{\bf{p}}^2} 
\end{eqnarray} and $M_H$ is the meson mass.  The meson and the
constituent quark states are normalized as \begin{eqnarray} \langle
M({\bf{P^\prime}_M},J^{\prime},m^{\prime}_J)|M({\bf {P}_M},J,m_J) \rangle\ &=&
2E\delta^3({\bf {P^\prime}_M}-{\bf{P}_M})\delta_{J^{\prime},J}\delta_{m^{\prime}_J,m_J}
\end{eqnarray} \begin{eqnarray} \langle
q({\bf{p^\prime}},m^{\prime}_s)|q({\bf{p}},m_s) \rangle\ &=&
{E_q\over{m_q}} \delta^3({\bf {p^\prime}}-{\bf
{p}})\delta_{m^{\prime}_s,m_s} \end{eqnarray}

In constructing the meson states we maintain a constituent quark
model approach as we do not include $q\bar{q}$ sea quark states nor the
explicit gluonic degrees of freedom. We also assume the validity
of the weak binding approximation \cite{ISGW1,ISGW2}. 
 In the weak binding limit our meson state forms a representation of the
Lorentz group, as discussed in Ref.\cite{ISGW1}, if the quark momenta
are small compared to their masses. Assuming that the quark fields in
the current create and annihilate the constituent quark states appearing
in the meson state, the calculation of the matrix element of the current operator then reduces to the calculation of a
free quark matrix element.  In the rest frame of the $B$ meson with a
suitable choice of the four-vector indices in Eqs.(6,7) we can construct
six independent equations which we can solve to extract the six form
factors.

This model space representation may be viewed as the leading
characterization in an expanded representation which more accurately
represents the exact states. We assume that the effects of Higher Fock
states, representing gluons or sea quarks, in the calculation of the matrix element of the bare current operator are
 represented by the matrix element of an effective operator in the model
space. In other words, with the notation ``e'' labelling exact states, and ``m'' labelling model states,
\ber
<M_2^e({\bf {P_2}})|J_{\mu}|M_1^e({\bf {P_1}})> & \to 
& <M_2^m({\bf {P_2}})|J_{\mu}^{eff}|M_1^m({\bf {P_1}})> \
\eer
where the higher Fock state effects are included by the following replacement in the calculation of the matrix element
\ber
\Phi_2^e({\bf {p'}})  J_{\mu}\Phi_1^e({\bf {p}}) 
& \to &
\Phi_2^m({\bf {p'}})  J_{\mu}^{eff}\Phi_1^m({\bf {p}}) 
=\Phi_2^m({\bf {p'}}){\Omega^{\dagger}({\bf {p'}})} J_{\mu}{\Omega({\bf
{p}})} \Phi_1^m({\bf{p}}) \
\eer
In the above, ${\bf p}$, ${\bf p'}$ 
are the internal momenta of the quarks in the initial and final mesons and 
$\Phi_1({\bf{p}})$, $\Phi_2({\bf{p'}})$ 
are the initial and final  meson wavefunctions.
We will use the very simple ansatz
\ber
{\Omega({\bf {p}})}
  & = & e^\frac{-\alpha^2 {{\bf {p^2}}}}{2}\
\eer

  We will
fix the parameter $\alpha$ by simply fitting to the available
experimental data and lattice results of the decay constants of the
leptonic decays. Note that we will use the same
value of $\alpha$ for decays involving B and D decays. This is consistent
with  Heavy Quark Symmetry. 

The expressions of the decay constants in terms of the wavefunctions are
given as \cite{VI}
\ber
f_i & = & \sqrt{\frac{12}{M}}
\int^{\infty}_{0}\frac{p^2dp}{2 \pi^3}
\sqrt{\frac{(m_q + E_q)(m_{\bar{q}} + E_{\bar q})}{4E_qE_{\bar{q}}}}
F_i(p)\\
F_{P}(p) & = & \left[ 1-\frac{p^2}{
(m_q + E_q)(m_{\bar{q}} + E_{\bar q})}\right]\psi_P(p)\\
F_V(p) & = & \left[ 1-\frac{p^2}{
3(m_q + E_q)(m_{\bar{q}} + E_{\bar q})}\right]\psi_V(p)\
\eer
where $\psi_{P(V)}$ are the wavefunctions of the exact states. Using Eq.(12), we can then obtain the form factor in terms of the BSE wavefunctions.
\section{ Non-leptonic Decays}
Non-leptonic decays arise from W exchange diagrams at tree level. Strong
interactions play an important role in these decays by  modifying
the weak vertices through hard gluon corrections and then the long
distance QCD interactions result in the binding of the quarks in the
hadrons. An effective Hamiltonian of four quark operators
 is constructed by integrating the
W-boson and the top quark from the theory. The effects of the short
distance and the long distance QCD interactions
are separated using the operator product expansion where the Wilson coefficients
account for the short distance effects while the long distance effects
are incorporated in the matrix element of the four quark operators. The
effective Hamiltonian operator for $b \to c $ transition can be written as
\ber
H_{eff} & = & \frac{G_F}{\sqrt 2} V_{cb} V_{ud}^*
            \left[ c_1(\mu)O_1 + c_2(\mu)O_2 \right ] \nonumber\\
O_1 & = & {\bar d_i}\gamma_{\mu}(1 -\gamma_5)u_i
                    {\bar c_j}\gamma^{\mu}(1 -\gamma_5)b_j \nonumber\\
O_2 & = & {\bar d_i}\gamma_{\mu}(1 -\gamma_5)u_j
                    {\bar c_j}\gamma^{\mu}(1 -\gamma_5)b_i \
\eer
where $i$ and $j$ are the color indices. The Wilson's coefficients $c_1$
and $c_2$ at the scale $\mu=m_b$ have values 1.132 and -0.286
respectively \cite{SN}.

The matrix element of a two body leptonic decay of the type $ B \to X Y
$ requires the evaluation of the matrix element
\bers
M & = & < X, Y|H_{eff}| B > \
\eers
where $H_{eff}$ has a current $\times$ current structure.
 The matrix element is usually calculated using the factorization
assumption where one separates out the current in  $H_{eff}$ by
inserting the vacuum state and neglecting any QCD interactions between
the currents. The matrix
element above written as a product of two current matrix elements is 
\ber
M & \sim & <X|J_{\mu}|0><Y|J^{\prime \mu}|B> \
\eer

In B decays, for e.g $ B \to D^+ \pi ^-$, the energetic quark-antiquark pair in
the pion is created at short distance 
 and by the time it hadronizes it
is far from the other quarks
 so it should be a good approximation to neglect the QCD interaction between the two
currents creating the final sate particles. A detailed description about
the validity and the corrections to the factorization assumption can be
found in Ref.\cite{SN}.

In this paper we will look at decays where the particle $Y$ is a $D$ or
a $D^*$ meson because one can then use the semi-leptonic form factors
calculated in the previous section to compute
$<Y|J^{\prime \mu}|B>$. The $X$ will be either a light meson(
$\pi$,$K$,$\rho$,$K^*$) or a $D(D^*)$ meson. For the light mesons the decay constants
$f_X=<X|J_{\mu}|0>$ are available from experiment while, for the heavy
mesons, we will use the decay constants calculated in the previous
section.

The expressions for the square of the matrix element for the processes
$ {\bar B}^0 \to D(D^*) \pi(\rho) $ are
\ber
|M|^2( {\bar B}^0 \to D^+ \pi^-) & = &
({\frac{G_F}{\sqrt 2}})^2 |V_{cb}V_{ud}^*|^2 (c_1 +c_2/N_c)^2
f_{\pi}^2 {F_0(m_{\pi}^2)}^2 (m_B^2-m_D^2)^2\
\eer
where $N_c$ is the number of colors.
\ber
|M|^2( {\bar B}^0 \to D^+ \rho^-) & = &
({{G_F}\sqrt 2})^2 |V_{cb}V_{ud}^*|^2 (c_1 +c_2/N_c)^2
m_{\rho}^2f_{\rho}^2 F_1(m_{\rho}^2)^2 m_B^2\frac{p^2}{m_{\rho}^2}\
\eer
 where $p$ is the momentum of the decay products in the rest frame of
the $B$.
\ber
|M|^2( {\bar B}^0 \to D^{+*} \pi^-) & = &
({\frac{G_F}{\sqrt 2}})^2 |V_{cb}V_{ud}^*|^2 (c_1 +c_2/N_c)^2
f_{\pi}^2 4m_{D^*}^2A_0(m_{\pi}^2)^2 m_B^2\frac{p^2}{m_{D^*}^2}\
\eer
and finally,
\ber
|M|^2( {\bar B}^0 \to D^{*+} \rho^-) & = &
({\frac{G_F}{\sqrt 2}})^2 |V_{cb}V_{ud}^*|^2 (c_1 +c_2/N_c)^2
m_{\rho}^2f_{\rho}^2 \left[ T_1 + T_2 + T_3 + T_4\right ]\nonumber\\
T_1 & = & \frac{8V^2}{(m_B + m_{D^*})^2} p^2m_B^2\nonumber\\
T_2 & = & A_1^2{(m_B + m_{D^*})^2}\left[
 2 +\frac{(m_BE_{D^*}-m_{D*}^2)^2}{m_{D*}^2m_{\rho}^2}
\right]\nonumber\\
T_3 &= & \frac{4A_2^2}{(m_B + m_{D^*})^2} 
\frac{p^4m_B^4}{m_{D*}^2m_{\rho}^2}\nonumber\\
T_4 &=&4A_1A_2\left[
\frac{E_{\rho}^2}{m_{\rho}^2} +\frac{E_{D^*}^2}{m_{D^*}^2}
-\frac{E_{D^*}E_{\rho}(m_BE_{D^*}-m_{D*}^2)}{m_{D*}^2m_{\rho}^2}
-1
\right]\
\eer
Similar expressions can also be written down for the $B \to D D $
decays.

As in Ref.\cite{SN} we will include the effect of corrections to the
factorization assumption by the replacement
\bers
c_1 +c_2/N_c & \to & a_1\\
a_1 & = & c_1(\mu) +\frac{c_2(\mu)}{N_c}\left (1
+\varepsilon^{1}(\mu)\right)
+c_2(\mu) \varepsilon^8(\mu)\
\eers
The nonfactorizable corrections $\varepsilon^1(\mu)$ and
$\varepsilon^8(\mu)$ are defined in Ref.\cite{SN} and may be process
dependent. We will, however, treat $a_1$ 
as a process independent free
parameter that we will fit to data. RGE analysis suggests that
$a_1 \sim 1 + 0(1/N_{c}^2)$.
\section{Results and Discussions}
 In previous papers \cite {BSV,BSV1}, a covariant reduction of the Bethe
-Salpeter equation (BSE) was used to calculate the Isgur-Wise function.
The BSE was solved numerically and the parameters appearing in it (the
quark masses, string tension and the running coupling strength for the
one gluon exchange) were determined by fitting the calculated spectrum
to the observed masses of more than 40 mesons.
The resulting mass spectrum of the analysis was found to agree
 very well with the experimental data. Once the parameters of the model
were fixed, the meson wavefunction could be calculated from the
BSE. This wavefunction was used to calculate the Isgur-Wise function and
determine $V_{cb}$ \cite{BSV}. 

In our present approach we evaluate the decay constants, the form
factors for the semileptonic decays ${\bar B}  \to D^* l {\bar{\nu}}  $ and ${\bar
B}  \to D l {\bar{\nu}}  $ with the effective current operator defined in Eq.(13) treating $\alpha$ of Eq.(14) as an
adjustable parameter. The value of $\alpha$ is fixed by fitting the
leptonic decay constants. We find $\alpha=0.7 GeV^{-1}$ provides a good fit and
use this value in all the calculations in
this paper.

We present our results for the decay constants of the heavy mesons
 in Table 2. For the sake of comparison we also show lattice calculations
of the decay constants.
\begin{table}
\caption{Decay constants of the B and D mesons in MeV}
\begin{center}
\begin{tabular}{|c|c|c|}
\hline
Decay Constants & Our Results & Lattice Results\cite{Lat} \\
\hline
$f_{D}$ & $209$  & $196(9)(14)(8)$ \\
\hline
$f_{D^{*}}$ & $237$  & $- -$ \\
\hline
$f_{D_s}$ & $213$  & $211(7)(25)(11)$ \\
\hline
$f_{D_s^{*}}$ & $242$  & $- -$ \\
\hline
$f_{B}$ & $155$  & $166(11)(28)(14)$ \\ 
\hline
$f_{B^*}$ & $164$  & $- -$ \\
\hline
\end{tabular}
\end{center}
\end{table}
The errors in the second column of the Table are, respectively, $(1)$ the
statistical errors; $(2)$ the systematic errors of changing fitting ranges, as well as
 other errors within
the quenched approximations; and $(3)$ the quenching error. The results in
Table 2 show that our calculated decay constants are similar to the
lattice results.

On the other hand, our calculation for the decay constants of the light
mesons $\pi, K$ etc are not in good agreement with the experimental
numbers. In fact,  the light meson decay constants are larger than
experiment by a
factor of $2$. This is not surprising as our formalism is designed for the
heavy meson system.

In Fig.1 we show the form factors $F_0, F_1,
V, A_0, A_1$, and $A_2$ as a function of $q^2$. In Fig.2 we show a plot of the differential decay rate
for ${\bar B}  \to D^* l {\bar{\nu}}$. We obtain a good agreement with
the shape of the experimental data \cite{CLEO} and extract $|V_{cb}|=0.039 \pm 0.002 $. This is within the
range of the presently accepted values for $|V_{cb}|$ \cite{PRD2}. 

For the decay ${\bar{B}} \to D l {\bar{\nu}}$, in Fig.3 we show a plot of $F(\omega)|V_{cb}|$ versus $\omega$ where the data points are taken from measurements reported in Ref.\cite{newcleo}. The variable $\omega = (M_B^2 +M_D^2 - q^2)/(2M_BM_D)$ where $q
^2$ is the invariant mass squared of the lepton neutrino system. We extract $|V_{cb}|=0.037 \pm 0.004$ by a $\chi^2$ fit to the data in Fig.3.

Note that the values of $V_{cb}$ extracted from the two different
experiments are consistent with each other. As a further test of our formalism we present our calculations of the
nonleptonic decays of the B meson to $DD$ and $DK(\pi)$ final states.
Experimental values of some of the decays are already available and new
results are expected soon. We present our results in Table 3. The
values of the light decay constants used in our calculations are
$f_{\pi}=130$MeV, $f_K=159$ MeV, $f_{K*}= 214$ MeV and $f_{\rho}=208 $ MeV.

\begin{table}
\caption{Non-Leptonic Decay Rates for B meson}
\begin{center}
\begin{tabular}{|c||c|c|c|}
\hline
Process & Our Results & Stech-Neubert \cite{SN}& Expt \cite{Expt}\\
\hline
${\bar{B^0}} \to D^{+} \pi^{-}$ & $0.345$  &$0.300$ & $0.310(0.040)(0.020)$ \\
\hline
${\bar{B^0}} \to D^{*+} \pi^{-}$ & $0.331$  &$0.290$ & $0.280(0.040)(0.010)$ \\
\hline
${\bar{B^0}} \to D^{+} \rho^{-}$ & $0.799$  &$0.750$ & $0.840(0.160)(0.070)$ \\
\hline
${\bar{B^0}} \to D^{*+} \rho^{-}$ & $0.897$  &$0.850$ &$0.730(0.150)(0.030)$\\
\hline
${\bar{B^0}} \to D^{+} K^{-}$ & $0.26$  &$0.20$ &$- -$ \\
\hline
${\bar{B^0}} \to D^{*+} K^{-}$ & $0.24$  &$0.20$ &$- -$ \\
\hline
${\bar{B^0}} \to D^{+} K^{*-}$ & $0.41$  &$0.40$ &$- -$ \\
\hline
${\bar{B^0}} \to D^{*+} K^{*-}$ & $0.49$  &$0.50$ &$- -$\\
\hline
${\bar{B^0}} \to D^{+} D^{-}$ & $0.31$  &$0.40$ &$- -$ \\
\hline
${\bar{B^0}} \to D^{*+} D^{-}$ & $0.22$  &$0.30$ &$- -$ \\
\hline
${\bar{B^0}} \to D^{+} D^{*-}$ & $0.27$  &$0.30$ &$- -$ \\
\hline
${\bar{B^0}} \to D^{*+} D^{*-}$ & $0.65$  &$0.80$ &$- -$\\
\hline
${\bar{B^0}} \to D^{+} D_s^{-}$ & $0.626$  &$1.030$ &$0.740(0.22)(0.18) $ \\
\hline
${\bar{B^0}} \to D^{*+} D_s^{-}$ & $0.420$  &$0.700$ &$0.94(0.24)(0.23)$ \\
\hline
${\bar{B^0}} \to D^{+} D_s^{*-}$ & $0.514$  &$0.950$ &$1.140(0.42)(0.28)$ \\
\hline
${\bar{B^0}} \to D^{*+} D_s^{*-}$ & $1.35$  &$2.450$ &$2.0(0.54)(0.05)$\\
\hline
${B^-} \to D^{0} D^{-}$ & $0.33$  &$0.40$ &$- - $ \\
\hline
${B^-} \to D^{*0} D^{-}$ & $0.210$  &$0.30$ &$- -$ \\
\hline
${B^-} \to D^{0} D^{*-}$ & $0.27$  &$0.40$ &$ - -$ \\
\hline
${B^-} \to D^{*0} D^{*-}$ & $0.64$  &$0.90$ &$ - -$\\
\hline
${B^-} \to D^{0} D_s^{-}$ & $0.829$  &$1.090$ &$ 1.360(0.280)(0.330) $ \\
\hline
${B^-} \to D^{*0} D_s^{-}$ & $0.552$  &$0.750$ &$0.940(0.310(0.23)$ \\
\hline
$B^- \to D^{0} D_s^{*-}$ & $0.696$  &$1.020$ &$ 1.180(0.36)(0.29)$ \\
\hline
$B^- \to D^{*0} D_s^{*-}$ & $1.830$  &$2.610$ &$ 2.700(0.810)(0.660)$\\
\hline
\end{tabular}
\end{center}
\end{table}

The parameter $a_1$ calculated on the basis of a $\chi^2$ fit has the
value $0.88$ which is close to 1 as is expected from RGE analysis which
gives $a_1 \sim 1 +O(1/N_c^2)$ suggesting a value for $a_1$ in the range $0.9-1.1$

>From Table 3 we find a good agreement of our calculation with
data for the $D^{*}\pi(\rho)$ final states. Our results for the $D K$
final states are quite similar to those in Ref.\cite{SN}. This continues
to be true for the $D D$ and $D^0 D^-$ final states. Our results for
$D^+D_s^-$
and $D^0D_s^-$
final states are somewhat smaller than the central values from
experiment though the measurements have large errors.

Combining the
experimental errors in quadrature the difference between theory and
experiment is less than $1.5 \sigma$ in all cases but the theory
predictions are systematically lower for these cases. It appears that as
we increase the mass of the decay products,
as in the $DD$ final states, and decrease their kinetic energy the
expected deterioration of the factorization approximation may be showing
up through a systematic difference between theory and experiment.
This motivates future efforts to examine corrections to the
factorization approximation \cite{SN}. We have resisted the temptation
to allow $a_1$ to have a process dependence even though two values for
$a_1$ would yield an excellent description of the known nonleptonic
decay rates. It is trivial to relax this restriction if the reader
chooses to do so.

	In conclusion, we have presented the calculation of form factors
and differential decay rates in  $ {\bar B}   \rightarrow D(D^*)l {\bar{\nu}} $ transitions in a Bethe-Salpeter model for mesons.
 The parameters of the bound state model were fixed from the spectroscopy of the hadrons.
The effects of higher Fock states in the hadron state were included in
the definition of effective current operators. A simple ansatz
connecting the effective current operator to the actual current
 operator was used involving
 only one parameter. After adjusting this parameter to fit certain decay
constants, we found good agreement with data and extracted
$|V_{cb}|=0.039 \pm 0.002$
from
${\bar B} \to D^* l {\bar{\nu}} $ and $V_{cb}=0.037 \pm 0.004$ from
${\bar B} \to D l {\bar{\nu}}  $  decays.
 Calculations of the decay constants of the B and
D mesons were also performed with results that are similar to lattice results. Finally, the form factors were used to evaluate the
non-leptonic $B \to D \pi(K)$ and $B\to D D(D_s)$ decays in the
factorization
approximation and good agreement was obtained with data.

{\bf Acknowledgments}
This work was supported in part by the US Department of Energy, Grant
No. DE-FG02-87ER40371, Division of High Energy and Nuclear Physics and Natural Sciences and Engineering Council of Canada.

%\newpage

\section{\bf Figure Captions}

\begin{itemize}

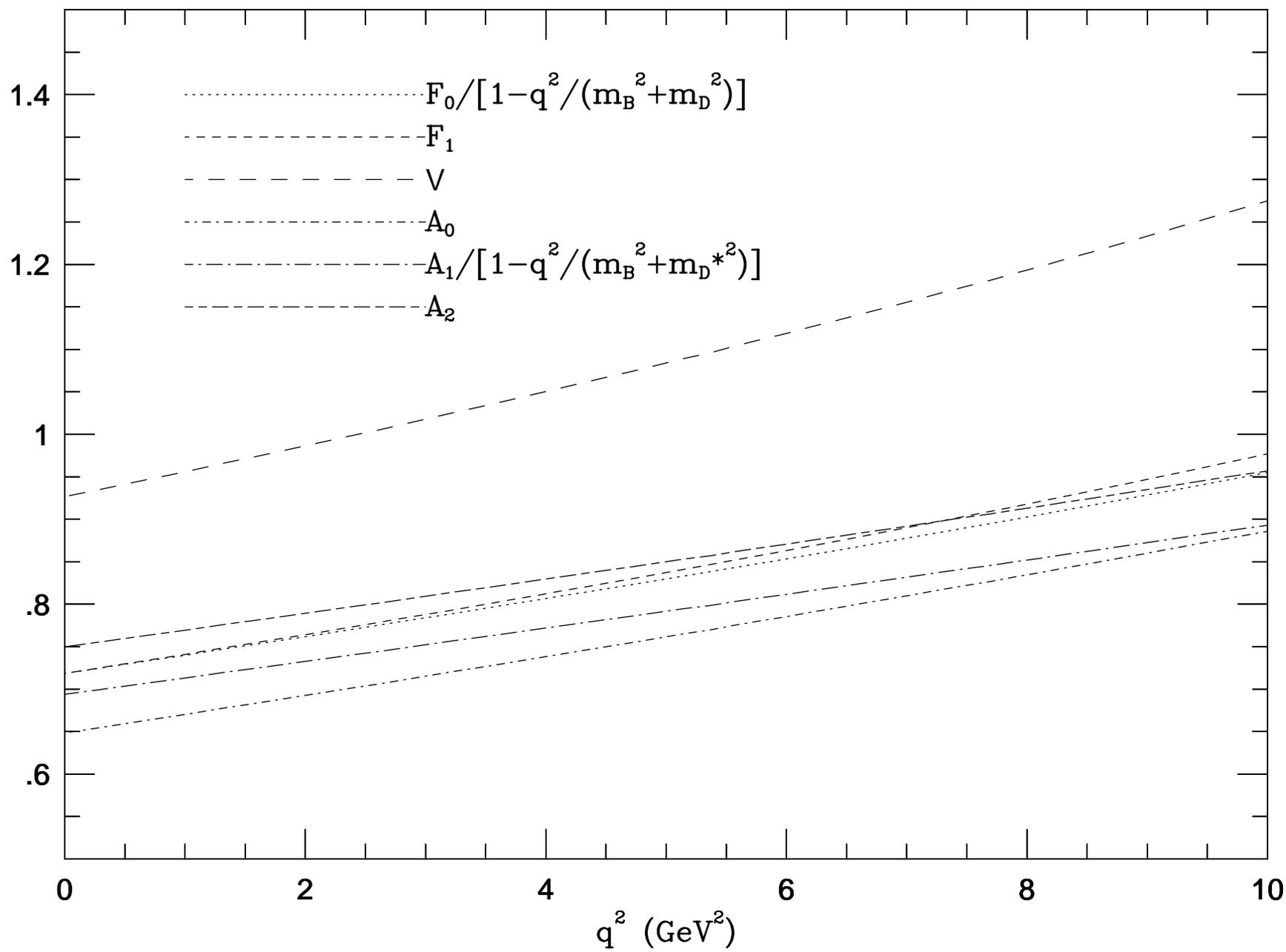
\item[\bf{Fig.1:}] The calculated form factors  $F_0, F_1,
V, A_0, A_1$, and $A_2$ as a function of $q^2$.

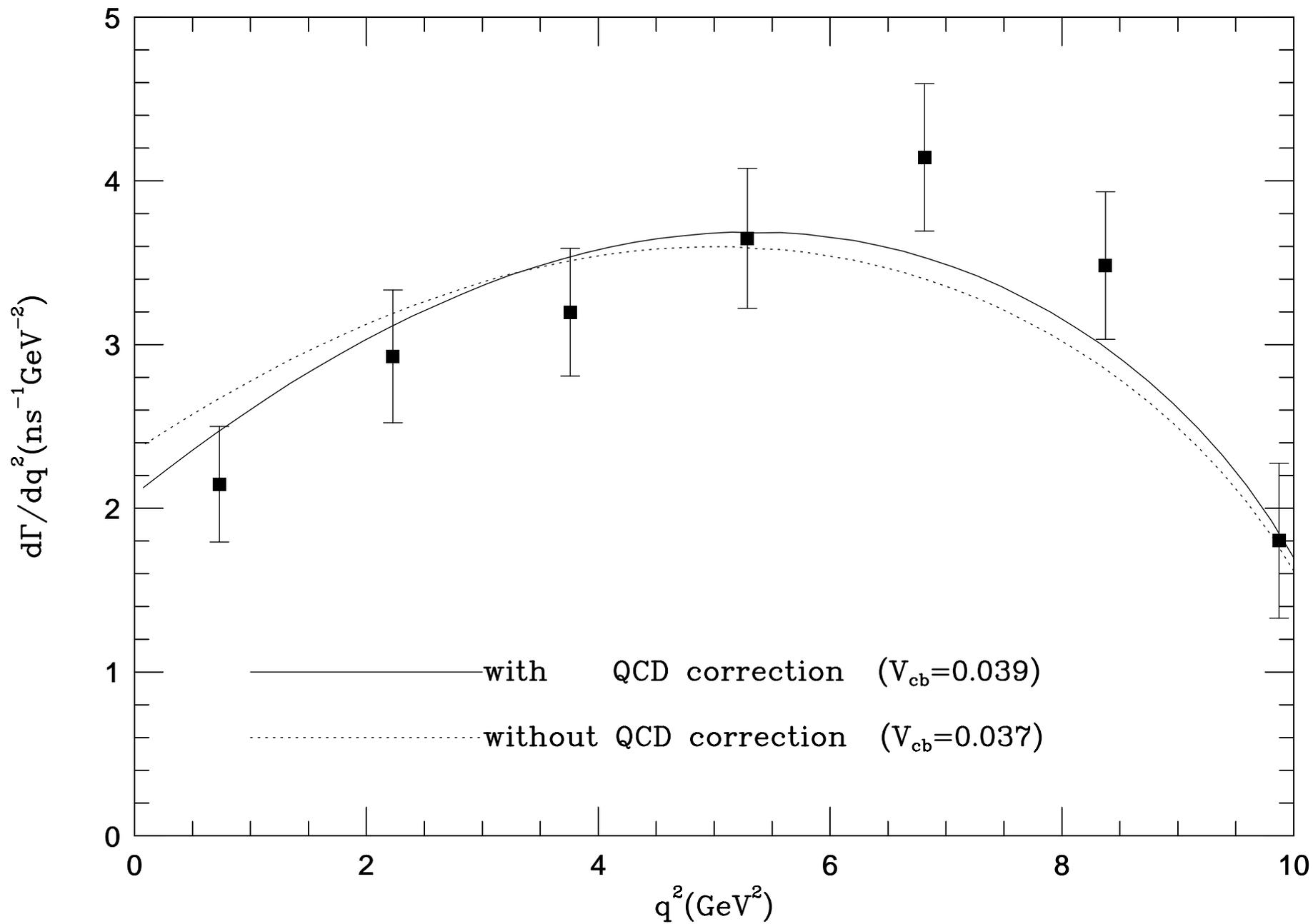
\item[\bf{Fig.2:}] The differential decay rate for ${\bar B}  \to D^* l{\bar{\nu}}   $
with and without the QCD correction, together with the corresponding
values of $V_{cb}$. Data from Ref.\cite{CLEO}.

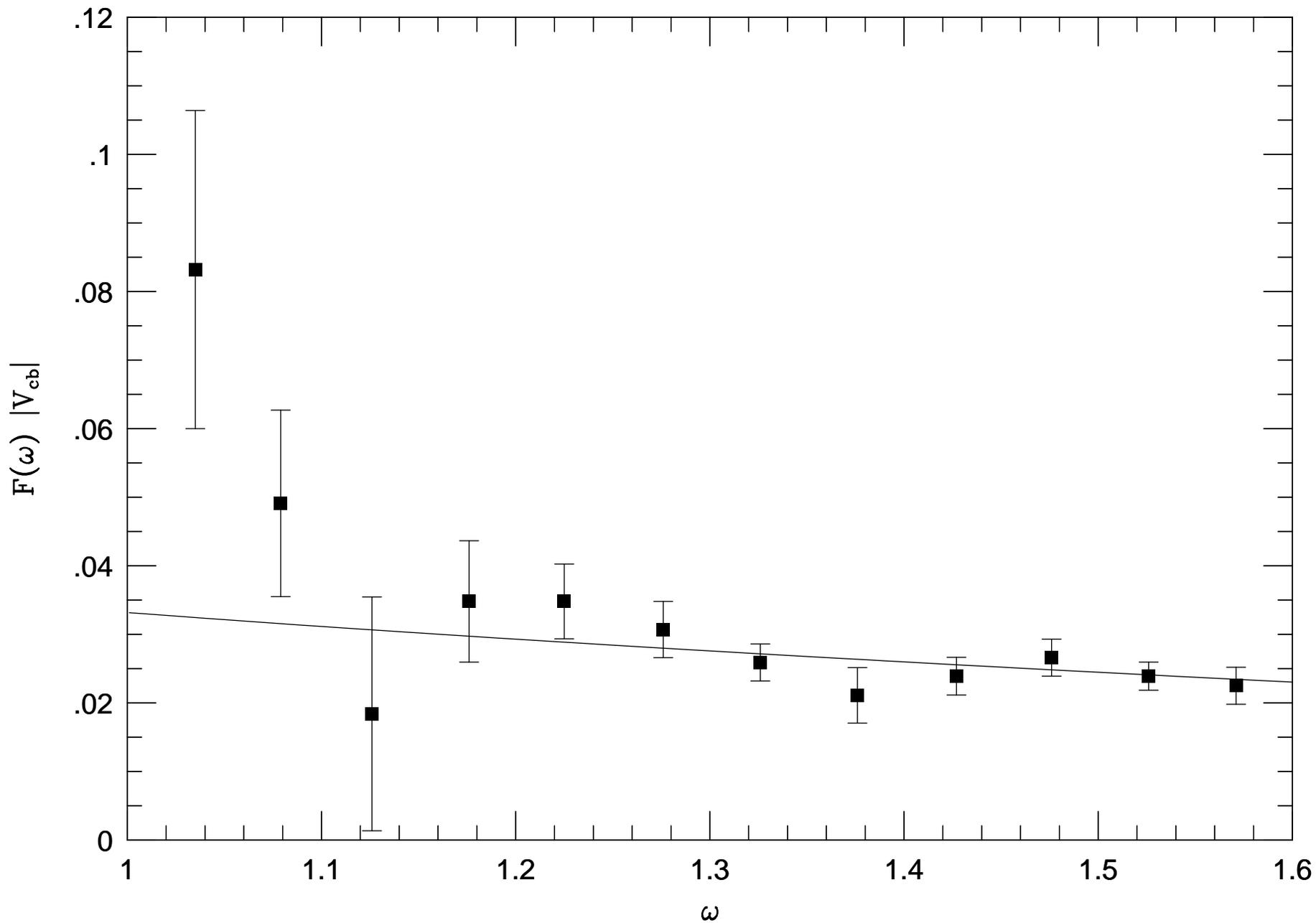
\item[\bf{Fig.3:}] $F(\omega)|V_{cb}|$ versus $\omega$ for ${\bar B} \to
D l {\bar{\nu}} $. Data from Ref.\cite{newcleo} 

\end{itemize}


\begin{thebibliography}{References}

\bibitem{BS}
D. Eyre and J.P. Vary, 
Phys. Rev. {\bf D 34}, 3467 (1986);
J.R. Spence and J.P. Vary, 
Phys. Rev. {\bf D35}, 2191 ( 1987);
J.R. Spence and  J.P. Vary, 
Phys. Rev. {\bf C47}, 1282 (1993);
A.J. Sommerer, J.R. Spence and J.P. Vary, 
Phys. Rev. {\bf C49}, 513 (1994).

\bibitem{BSsep}
Alan J. Sommerer, A. Abd El-Hady, John R. Spence, and James P. Vary, 
Phys. Lett. {\bf B348}, 277 (1995). 

\bibitem{BSV}
A. Abd El-Hady, K.S. Gupta, A.J. Sommerer, J. Spence, and  J.P. Vary, 
Phys. Rev. {\bf D51}, 5245 (1995). 

\bibitem{BSV1}
A. Abd El-Hady, A. Datta, K.S. Gupta, and  J.P. Vary, 
Phys. Rev. {\bf D51}, 5245 (1997). 

 \bibitem{H4} N. Isgur and M.B. Wise, 
Phys. Lett. {\bf B232}, 113 (1989); 
Phys. Lett. {\bf B237}, 527 (1990);
N. Isgur and M.B. Wise, 
``Heavy Quark Symmetry" in {\it B Decays}, 
ed. S. Stone (World Scientific, Singapore, 1991), p. 158,
{\it ``Heavy Flavors"}, ed. A.J. Buras and M. Lindner 
(World Scientific, Singapore, 1992), p. 234.

\bibitem{H5}
M. B.~Voloshin and M. A. Shifman, Yad. Fiz. {\bf 47}, 801 (1988); 
Sov. J. Nucl. Phys. {\bf 47}, 511 (1988); 
M. A. Shifman in {\em Proceedings of the 1987 International Symposium 
on Lepton and Photon Interactions at High Energies}, Hamburg, 
West Germany, 1987, edited by W. Bartel and R. R{\"u}ckl, 
Nucl. Phys. B (Proc.. Suppl.) {\bf 3}, 289 (1988); 
S. Nussinov and W. Wetzel, Phys. Rev. {\bf D36}, 130 (1987);
G.P. Lepage and B.A. Thacker, in {\it Field Theory on the Lattice}, 
edited by A.~Billoire, Nucl. Phys. B (Proc. Suppl.) 4 (1988) 199;
E. Eichten, in {\it Field Theory on the Lattice}, 
edited by A.~Billoire, Nucl. Phys. B (Proc. Suppl.) 4, 170 (1988);
E. Shuryak, Phys. Lett. {\bf B93}, 134 (1980); 
Nucl. Phys. {\bf B198}, 83 (1982).

\bibitem{H6}
H. Georgi, Phys. Lett. {\bf B240}, 447 (1990); 
E. Eichten and B. Hill, Phys. Lett. {\bf B234}, 511 (1990); 
M. B. Voloshin and M.A. Shifman, Sov. J. Nucl. Phys. {\bf 45}, 463 (1987);
H.D. Politzer and M.B. Wise, Phys. Lett. {\bf B206}, 681 (1988); 
Phys. Lett. {\bf B208}, 504 (1988); 
A.F. Falk, H. Georgi, B. Grinstein and M.B. Wise,  
Nucl. Phys. {\bf B343}, 1 (1990);
B. Grinstein,  Nucl. Phys. {\bf B339}, 253 (1990); 
M.B. Wise, ``CP Violation" in {\it Particles and Fields 3:
Proceedings of the Banff Summer Institute (CAP)} 1988, p.~124, 
edited by N.Kamal and F. Khanna, World Scientific
(1989). 

\bibitem{H7}
C.O. Dib and F. Vera, Phys. Rev. {\bf D47}, 3938 (1993);
J.F. Amundson, Phys. Rev. {\bf D49}, 373 (1994);
J.F. Amundson and J.L. Rosner, Phys. Rev. {\bf D47}, 1951 (1993);
B. Holdom and M. Sutherland, Phys. Rev. {\bf D47}, 5067 (1993).

\bibitem{H71}
E. Bagan, P. Ball, V.M. Braun, and H.G. Dosch, 
Phys. Lett. {\bf B278}, 457 (1992); 
M. Neubert, Phys. Rev. {\bf D46}, 3914 (1993); 
M. Neubert, Z. Ligeti, and Y. Nir, Phys. Lett. {\bf B301}, 101 (1993); 
Phys. Rev. {\bf D47}, 5060 (1993).

\bibitem{newcleo} M. Athanas {\it{et al.,}} Phys. Rev. Lett. {\bf 79}
2208 (1997); We have extracted the data and errors from the published figure.

\bibitem{BSW}
M. Wirbel, B. Stech, and M. Bauer, Z. Phys. {\bf C29}, 637 (1985); M.
Bauer, B. Stech, and M. Wirbel Z. Phys. {\bf C34}, 103 (1987).

\bibitem{NR}
M. Neubert, Phys. Rep. {\bf 245}, 259 (1994) and references therein;
M. Neubert, Int. J. Mod. Phys. {\bf A11}, 4173 (1996).

\bibitem{ISGW1}
Nathan Isgur, Daryl Scora, Benjamin Grinstein, and Mark B. Wise, Phys. Rev. 
{\bf D 39}, 799 (1989).

\bibitem{ISGW2}
Daryl Scora and Nathan Isgur, 
Phys. Rev. {\bf D52}, 2783 (1995).

\bibitem{VI} S. Veseli and I. Dunietz Phys. Rev. {\bf D 54} 6803 (1996). 

\bibitem{SN} M, Neubert and B. Stech, hep-ph/9705292; to appear in the
second Edition of Heavy Flavours, edited by A. J. Buras and M. Linder
(World Sceintific, Singapore).

\bibitem{CLEO}
B. Barish {\it et al.}, (CLEO)
Phys. Rev. {\bf D51}, 1014 (1995); J.E. Duboscq {\it et al.}, (CLEO)
Phys. Rev. lett.{\bf 76}, 3898 (1996).

\bibitem{PRD2}
Particle Data Group, R.M. Barnett {\it et al.,}, Phys. Rev. {\bf D54}, 1 (1996).  

\bibitem{Lat} C. Bernard {\it {eta.al}} Nucl. Phys. Proc. Suppl. {\bf
53}, 358 (1997).
  
\bibitem{Expt} J. D. Rodriguez, to appear in : Proceedings of the 2nd
International Conference on B Physics and CP Violations, Honolulu,
Hawaii, March 1997; T. E. Browder, K. Honscheid and D. Pedrini, Ann.
Rev. Nucl. Part. Sci. {\bf 46}, 395 (1996).

\end{thebibliography}
\end{document}